\documentclass[11pt, fleqn]{article}
\usepackage{amsmath}
\usepackage{amssymb,amsthm,epsfig}
\usepackage{booktabs, multirow} 
\usepackage{changepage,threeparttable} 
\usepackage{enumerate}
\usepackage{mathabx}
\usepackage{tcolorbox}
 \def\vt{t\kern-0.22em\raise.18ex\hbox{\char'47}\lower.18ex\hbox{}\kern-0.08em}
\newtheorem{theorem}{Theorem}[section]

\newtheorem{lm}{Lemma}[section]
\newtheorem{ex}{Example}[section]

\newtheorem{rem}{Remark}[section]

\newcommand{\old}[1]{{}} 

\usepackage{accents}
\newlength{\dhatheight}

\newcounter{obr}
\newcounter{tabul}
\begin{document}
\title{ A performance study of some approximation  algorithms  for minimum   dominating set in a graph 
}
\author{Jonathan S. Li$^1$, Rohan Potru$^1$, Farhad Shahrokhi$^2$}

\date{%
    $^1$The University of Texas at Austin\\%
    $^2$University of North Texas\\[2ex] %
    jonathansli@utexas.edu\\%
    rohanpotru@utexas.edu\\%
    farhad.shahrokhi@unt.edu\\%
}

\maketitle


\begin{center}\textit{$^1$Work done at UNT and supported by the Texas Academy of Math and Science.}\end{center}
\begin{abstract}
We implement and test the performances of several approximation algorithms for computing the minimum dominating set of a graph. These algorithms are the standard greedy algorithm,  the recent LP rounding algorithms  and  a hybrid algorithm that we design by combining the  greedy  and LP rounding algorithms. 
All algorithms  perform better than anticipated in their theoretical analysis, and  have small  performance ratios,   measured as the size of output divided by the LP objective lower-bound. However, each may have advantages over the others. For instance, LP rounding algorithm  normally outperforms the other algorithms on sparse real-world graphs. On a graph with 400,000+ vertices, LP rounding took less than 15 seconds of CPU time to generate a solution with performance ratio 1.011, while the greedy and hybrid algorithms generated solutions of performance ratio 1.12 in similar time. For synthetic graphs, the hybrid algorithm normally outperforms the others, whereas for hypercubes and k-Queens graphs, greedy outperforms the rest. Another  advantage of the hybrid algorithm is  to  solve very large problems where LP solvers  crash, as we observed on  a real-world graph with 7.7 million+ vertices.

\end{abstract}

\section{Introduction and Summary}
  Domination theory has its roots in the $k$-Queens problem in 18th century.  Later in 1957, Berge \cite{Be} formally introduced the domination number of a graph. The problem of computing the domination number of a graph has extensive applications including the design of telecommunication networks, facility location, and social networks. 
  We refer the reader to the book by Haynes, Hedetniemi,  and  Slater \cite{HHS} as a general reference in domination theory. \\
  
  We assume that the reader is  familiar  with general concepts  of  graph theory as in \cite{CL}, the theory of algorithms as in \cite{CLR}, and linear and integer programming concepts as in  \cite{AS}, respectively. 
  Throughout this paper $G=(V,E)$ denotes an undirected  graph  on vertex set $V$ and edge set $E$ with  $n=|V|$  and  $m=|E|$.  Two vertices $x,y \in V$ where $x\ne y$ are adjacent (or they are neighbors)  if $x,y\in E$. For any $x\in V$, degree of 
$x$, denoted by $deg(x)$ is  the number of vertices adjacent to $x$ in $G$. For any   $x\in V$, let $N(x)$ denote the set of all vertices in $G$ that are adjacent to $x$.  Let $N[x]$ denote $N(x)\cup \{x\}$. Arboricity of $G$, denoted by $a(G)$ is the minimum number of spanning acyclic subgraphs of $G$ 
that $E$ can be partitioned into. By a theorem of Nash Williams, $a(G)={\max_S}{\lceil {m_S\over n_S-1}\rceil}$, where $n_S$ and $m_S$ are the number of vertices and edges, respectively,  of the induced subgraph on the vertex set $S$ \cite{Nash}. Consequently $m\le a(G)(n-1)$, and thus $a(G)$  measures how dense $G$ is. It is known that $a(G)$ can be computed in polynomial time \cite{Ga}. \\
Let $D\subseteq V$. $D$ is a dominating set if for every $x\in V\setminus D$ there exists $y\in D$ such that $(x,y)\in E$. 
The {\it domination number} of $G$, denoted by $\gamma(G)$, is the cardinality of a  minimum (smallest) dominating set of $G$. Computing $\gamma(G)$ is known to be an NP-Hard problem  even for  unit disc graphs and grids \cite{CC}. 

\subsection{Greedy approximation algorithm}

A  simple greedy algorithm attributed to  Chvatal \cite{Ch} and Lovas \cite{Lo}  (for approximating the set cover problem) 
 is  known to approximate   $\gamma(G)$  within a multiplicative factor of $H(\Delta(G))$ from its optimal value, where $\Delta(G)$ is maximum degree of $G$ and $H(k)=\sum_{i=1}^k(1/i)$ is the $k-$th harmonic number.\footnote{Note that $ln(k+1)\le H(k)\le  ln(x)+1$.} The algorithm initially labels  all vertices uncovered. At iteration one, the algorithm  selects a  vertex $v_1$  of  maximum degree in $G$, places $v_1$ in a set $D$, and labels  all vertices adjacent to it  as  covered. In
 general,  at iteration $i\ge 2$, the algorithm selects a vertex $v_i\in  V-\{v_1,v_2,...,v_{i-1}\}$ with the largest number of  uncovered vertices adjacent to it , adds $v_i$ to $D$,  and labels   all of its uncovered  adjacent vertices as covered. The algorithm stops when $D$ becomes a dominating set. it is easy to implement the algorithm  in $O(n+m)$ time. 
 It is  known that   approximating $\gamma(G)$ within a factor $(1-\epsilon)ln(\Delta)$ from the optimal is NP-hard \cite{DD}. Hence, no  algorithm for approximation $\gamma(G)$    can  improve the asymptotic  worse case  performance ratio achieved  by the greedy algorithm.   Different variations of the greedy algorithm to approximate $\gamma(G)$ are developed and some are tested in practice; See work of  Chalupa \cite{Ch} Campan et. al. \cite{Cam}, Eubank et. al \cite{Eu}, Parekh \cite{Par}, Sanchis \cite{Sa}, and Siebertz \cite{Si}. \newline
 
Below are two examples of worst-case graphs (one sparse and one dense) for greedy algorithm which are derived from an instance of set cover problem provided in \cite{BG}.  For both instances, the solutions provided by the greedy algorithm are  actually $O(ln(\Delta))$ times the optimal. 
   
\begin{ex}\label{e1}
{\sl 
} 
\end{ex}

Let $p\ge 2$ be an integer and  for $i=1,2,...,p$,    let $S_i$ be a star on $2^{i}$ vertices.  Consider a graph $G$ on 
$n=2^{p+1}$ vertices whose vertices  are the disjoint union of the vertices of the $S_i$'s ($i=1,2,...,p$) plus two additional  vertices $t_1$ and $t_2$. Now, place edges from $t_1$ and $t_2$ to the first half of the vertices in each $S_i$  (including the root), and the second half of the vertices in each $S_i$, respectively.  Note that the root of each $S_i$ has degree  $2^{i}$ and the degree of both $t_1$ and $t_2$  is $2^{p}-1$. 
 Initially, greedy chooses the root of $S_p$ which can cover $2^p+1$ vertices (including itself).
 Generally, at iteration $i\ge 2$, there is a tie between the root of $S_{p+1-i}$ and $t_2$ since each can cover $2^{p-2}$ uncovered vertices. If tie breaking does not result in selecting $t_2$, there will be a tie in
 every iteration until the algorithm returns the set of $S_i$'s ($i=1,2,...,p$). This dominating set has cardinality $p=log(\Delta)-1$,  but 
$\gamma(G)=2$, since $\{t_1,t_2\}$ is a minimum dominating set. Note that $G$ is a planar graph.\\

\begin{center}\includegraphics[width=0.67\textwidth, height=0.23\textheight]{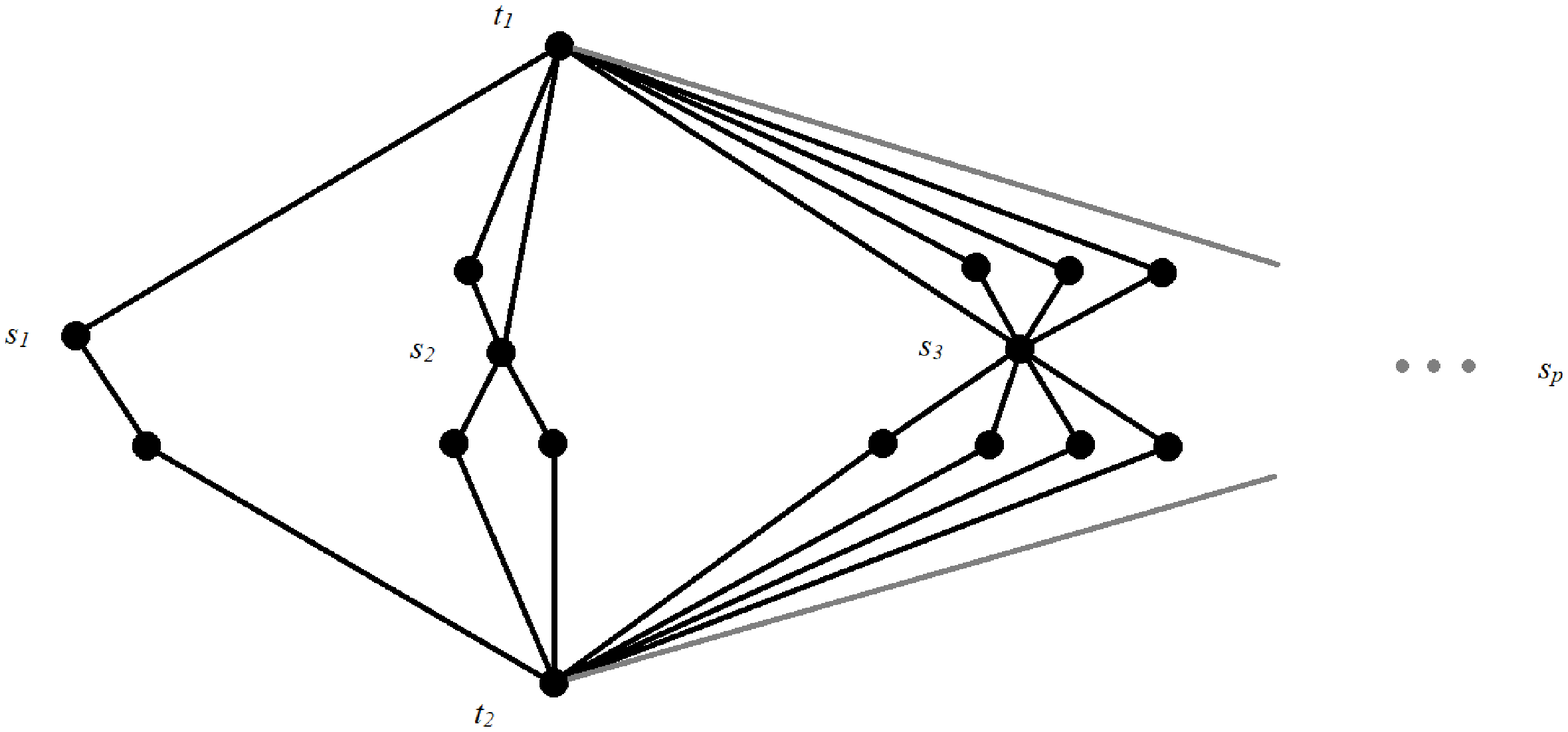}\end{center}

\begin{ex}\label{e2} 
{\sl }
\end{ex}
Let  $p\ge 2$ be an integer, and let $G$ be a graph with vertices $V_1\cup V_2$,where $V_1=\{s_1,s_2,...,s_p, t_1,t_2\}$ and $V_2=\{v_1,v_2,...,v_{2^{p+1}-2}\}$.  Now make $V_1$ a clique and $V_2$  an independent set of vertices, respectively. 
Next, consider a linear ordering $L$ on $V_2$: for $i=1,2,...,p$, the set of neighbors  of  $s_i$ in $V_2$, denoted by $W_i$,  has cardinality $2^i$ and is disjoint from   $W_k$, for any $k\le i$. Finally, for
$i=1,2,...,p$ place edges between $t_1$ and the first half of the vertices in each $W_i$, and place edges between $t_2$ and the second half of the vertices in  each $W_i$. Now note that the greedy algorithm will be
forced to pick the vertices  $s_p,s_{p-1},...,s_1$, in that order but  the minimum dominating set in $G$ is $\{t_1,t_2\}$ and $\Delta=2^p+p+1$. \\

 \subsection{Linear programing rounding approximation algorithms}
One can formulate the computation of  $\gamma(G)$ as an integer programming problem  stated below. However, since integer programming problems are known to be NP-hard \cite{Karp}, the
direct applications of the integer programming  method would not be computationally fruitful.

\begin{tcolorbox}[colback=white, arc=0mm]
\textbf{IP1:}\\
Minimize~~ $I = \sum_{v\in V} x_v$
\\
Subject~to~~$\sum_{u\in N[v]} x_u \geq 1,  \forall  v\in V$
 \\
$x_v \in \{0,1\}, ~~    \forall  v\in V$
\end{tcolorbox}

Now observe   that by relaxing the  integer program IP1 one obtains the following  linear program. 
\begin{tcolorbox}[colback=white,arc=0mm]
\textbf{LP1:}\\
Minimize~~$L = \sum_{v\in V} x_v$
\\
Subject~to~~$\sum_{u\in N[v]} x_u \geq 1,  \forall  v\in V$
 \\
~~$0\le x_v\le 1, ~~    \forall  v\in V$
\end{tcolorbox}

\vskip .3cm 
Note that  $L^*\le \gamma(G)= I^*$, where $L^*$ and $I^*$ are the values of $L$  and $I$ at optimality.  
 Since the  class of linear programming problems are solvable in polynomial time \cite{Kha},  LP1 can be solved in polynomial time. Very recently, Bansal and Umboh \cite{Ban} and Dvořák \cite{D2} have shown  that an
 appropriate rounding of  fractional solutions of LP1  gives integer solutions to IP1 whose values are at most  $3\cdot a(G)\cdot L^*$ and $(2\cdot a(G)+1)\cdot L^*$, respectively, in polynomial time. Hence, for sparse graphs (graphs with bounded arboricity), one can get a better approximation ratio than $O(ln(\Delta))$
 which is  achieved by the greedy algorithm. To our knowledge, and in contrast to the greedy algorithm, the performances of the LP rounding approaches  have  not been tested in practice.

\subsection{Other approximation algorithms}  There are other approximation algorithms for very specific  classes of graphs including planar graphs which have better than constant performance  ratio in the worst case
but are  more complex than algorithms described here. See \cite{Si}
 for a brief reference to some related papers. 
 


\subsection{Our work}
Greedy is simple and fast, since it can be implemented in linear time. Its performance ratio in the worst case scenario is logarithmic. Linear programming works in polynomial time but is more time consuming than greedy. For sparse graphs, recent linear programming rounding methods in \cite{Ban,D2} have a constant performance ratio, but  there have not been any experimental study of their performances. \newline

In this paper, we implement three types of algorithms and compare and contrast their performances in practice. These algorithms are  the greedy algorithm,   the LP rounding algorithms, and  a hybrid algorithm that combines the greedy and  LP approach.  The hybrid algorithm first  solves the problem using the greedy 
algorithm and finds a dominating set $D, |D|=d$. It then takes a portion of vertices in $D$, forces their weights  to be $1$ in linear program LP1, solves the resulting (partial) linear program, and then properly rounds the solution to the partial LP. Finally, it returns the rounded solution plus the portion of the greedy
solution that was forced to LP1. 
 

\subsection{ Environment,  implementation and datasets}
We used a laptop with modest computational power - 8th generation Intel i5 (1.6GHz) and 8GB RAM - to perform the experiments. We implemented the $O(n+m)$ time version of the greedy algorithm in C++.  We used  IBM Decision Optimization CPLEX Modeling (DOCPLEX) for Python to solve the LP relaxation of the problem. Python and DOCPLEX were used to implement the LP rounding and hybrid algorithms. \newline

The graph generator at $\dagger$ was used to create the planar graphs, trees, k-planar graphs (graphs embedded in the plane with at most $k$ crossings per edge) , and k-trees (graphs with tree width $k$ with largest number of edges) up to 20,000 vertices. The k-Queens graphs, hypercubes (up to 12 dimensions) and graph
implementations of the cases described in 1.1 and 1.2 were created ourselves. We also used publicly available Google+ and Pokec social-network graphs, as well as real-world DIMACS Graphs with up to more than 7,700,000 vertices. \newline

https://snap.stanford.edu/data/com-Youtube.html \cite{Cam}

https://github.com/joklawitter/GraphGenerators $\dagger$



http://davidchalupa.github.io/research/data/social.html \cite{Ch}

https://www.cc.gatech.edu/dimacs10/downloads.shtml \cite{Bader}

\subsection{Our results} 
Through experimentation, all algorithms  perform better than anticipated in their theoretical analysis, particularly with respect to the  performance ratios  (measured  with respect to the LP objective lower-bound).  However, each may have advantages over the others for specific data sets. For instance, LP rounding normally outperforms the other algorithms on real-world graphs. On a graph with 400,000+ vertices, LP rounding took less than 15 seconds of CPU time to generate a solution with performance ratio 1.011, while the greedy and hybrid
algorithms generated solutions of performance ratio 1.12 in similar time. For synthetic graphs (generated k-trees, k-planar)  the hybrid algorithm normally outperforms the others, whereas for hypercubes and k-Queens graphs, the greedy outperforms the rest. Particularly, on the 12-dimensional hypercube, greedy finds a solution with performance ratio 1.7 in 0.01 seconds. On the other hand, the LP rounding and hybrid algorithms produce solutions with performance ratio 13 and 3.3 using 7.5 and 0.08 seconds of CPU time, respectively. It is notable
that greedy gives optimal results in some cases where the domination number is known. 
 Specifically, the greedy algorithm produces an optimal solution on hypercubes with dimensions $d=2^k -1$ where k=1, 2, 3, and 4.

The hybrid algorithm can solve very large problems when the size of LP1 becomes formidable in practice.  
For instance, the  hybrid algorithm solved a real-world graph with 7.7 million+ vertices  in 106 seconds of CPU time with a performance ratio of 2.0075. The LP solver crashed on this problem. \newline

This paper is organized as follows. In section two, we formally describe LP rounding and hybrid algorithms. 
When the size of problem is so large that LP1 can not be solved in practice,  then  $L^*$ can not be computed, and hence the performance ratio of the hybrid algorithm can not be determined.  We resolved  this problem by decomposing LP1 in to two smaller linear programs so each of them has an
objective value not exceeding $L^*$ and used  the maximum objective value of the two smaller LP's, instead  of $L^*$, to measure the performance ratio of the hybrid algorithm. 
Section 3, 4, and 5 contains results for  Planar, k-Planar, and k-Tree graphs, hypercubes and k-Queen graphs, and real-world graphs respectively.



\section{Linear Programming and hybrid  approach}  

The following algorithm is due to  Bansal and Umboh \cite{Ban}.

\begin{tcolorbox}[colback=white,arc=0mm]
\textbf{Algorithm  $A_1$ (\cite{Ban})}
 
Solve  LP1, and  let $H$ be the set of all vertices that have weight at least $1/(3a(G))$, where $a(G)$ is the arboricity of graph $G$. Let $U$ be the set of all vertices not adjacent to any vertex in  $H$ and returns $H\cup U$. 
\end{tcolorbox}
Dvořák\cite{D1, D2} studied $d$-domination problem, that is, when a vertex  dominates all vertices at distance at most $d$ from it and its combinatorial dual, or  a $2d$-independent set \cite{BM}.  In \cite{D2} he  employed  the LP rounding approach of Bansal and Umboh, as a part of his  frame work  and  consequently, for $d=1$,  he improved  the approximation ratio of Algorithm $A_1$  by showing that the  algorithm  $A_2$ given below provides a   $2a(G)+1$ approximation. \\

\begin{tcolorbox}[colback=white,arc=0mm]
\textbf{Algorithm $A_2$  (\cite{D2})}

Solve  LP1, and   let $H$ be the set of all vertices that have weight at least $1/(2a(G)+1)$, where $a(G)$ is the arboricity of graph $G$.  Let $U$ be the set of all vertices that are not adjacent to any vertex of $H$ and return  $H\cup U$.

\end{tcolorbox}

\begin{rem}\label{R1}
 {\sl Graph $G$  in example \ref{e1} is planar, so  $a(G)\le 3$. Thus, algorithms $A_1$ and $A_2$  have a worst-case performance ratio of nine and seven respectively, whereas greedy exhibits a worst-case  $O(\log(n))$ performance ratio.  
 Throughout our experiments, rounding algorithms returned an optimal solution of size two for both examples, whereas greedy returned a set of size three for Example \ref{e1}. 
 Furthermore, in  Example \ref{e2}, it can be verified   that $a(G)\ge (p+2)/2$  for  graph $G$ and hence in theory the worse case performance ratios of the rounding algorithms are not constant either. Interestingly enough, in our experiments, $L^*$ was always two  for graphs of type Example \ref{e2}, and LP rounding algorithms  also always  found a solution of size two  which is the optimal value. 
 Thus the  performance ratio was always one  and much smaller than 
 the predicted worst case.
}
\end{rem}

Next, we provide a  description of the decomposition approach for  approximating  LP1 and our  hybrid algorithm. 
Recall that a  separation in  $G=(V,E)$ is  a partition  $A\cup B\cup C$ of  $V$ so that no vertex of $A$ is adjacent to any vertex of $C$. In this case  $B$ is called a  vertex separator in $G$.  Let $X=\{x_v|v\in V\}$ be a feasible solution to
LP1,  and let  $V'\subseteq V$. Then $X(V')$   denotes $\sum_{v\in V'} x_v$. 

\begin{lm}\label{l1}
{\sl Let  $A\cup B\cup C$ be a separation in $G=(V,E)$ and  consider the following linear programs:
\begin{tcolorbox}[colback=white,arc=0mm]
\textbf{LP2:}\\
Minimize~~   $M= \sum_{v\in {A\cup B} } x_v$
\\
Subject~to $\sum_{u\in N[v]} x_u \geq 1,  \forall  v\in A$
 \\
$0\le x_v\le 1, ~~    \forall  v\in A\cup B$
\end{tcolorbox}
and 
\begin{tcolorbox}[colback=white,arc=0mm]
\textbf{LP3:}\\
Minimize~~   $N= \sum_{v\in {C\cup B} } x_v$
\\
Subject~to $\sum_{u\in N[v]} x_u \geq 1,  \forall  v\in C$
 \\
$0\le x_v\le 1, ~~    \forall  v\in B\cup C$
\end{tcolorbox}

Then   
$\max\{ M^*, N^*\} \le L^*$.
}
\end{lm}
{\bf Proof.} 
Let $X=\{x_v|v\in V\}$ be an optimal solution to LP1. Note that the restrictions of $X$  to $A\cup B$ and $C\cup B$ give  feasible solutions for LP3 and LP2 of values $X(B\cup C)$ and $X(B\cup A)$, and hence the claim for the lower bound on $L^*$ follows.  $\Box$.

Note that in LP2, LP3  the constraints  are not written for all  variables, and rounding  method in \cite{Ban} may not directly be applied. 
  
\begin{theorem}\label{t1} 
{\sl Let  $G=(V,E)$, let $A\subset V$, let $B=E(A)$ and  let $C=V-(A\cup B)$. Let $X$ be an optimal solution for LP3, and let $X(C)$ denote the sum of the weights assigned to all vertices in $C$.
Then there is dominating set in $G$ of size at most $|A|+ 3a(G)X(C)\le |A|+3a(G)N^*$. 
}
\end{theorem}
{\bf Proof.} Let $H$ be the set of all vertices $v$  in $C$ with  $x(v)\ge {1\over 3a}$, and let $U=C-(H\cup E(H))$. Now apply the method in \cite{Ban} to $C$ to obtain a rounded solution, or a dominating set $D$,
of  at most $|U|+|H|\le 3a(G)X(C)$  vertices in  $C$. Finally, note that $A \cup D$ is a dominating set in $G$ with cardinality at most $|A|+ 3a(G)X(C)\le |A|+3a(G)N^*$ 
$\Box$. 

\begin{tcolorbox}[colback=white,arc=0mm]

\textbf{Algorithm $H$ (Hybrid Algorithm)}

Apply the greedy algorithm to $G$ to obtain a dominating set  $D=\{x1,x_2,....x_d\}$, and let $S=\{x_1,x_2,...,x_{\alpha.d}\}$ be the first $\alpha.d$ vertices in $D$.   Now solve the following linear program on the induced subgraph of $G$ with the vertex set $V-\{S\}$. 
 
\begin{eqnarray}
Minimize~~   J&= & \sum_{v\in V-\{S\}} x_v
\\
Subject~to & &\sum_{u\in N[v]} x_u \geq 1,  \forall  v\in V-\{S\cup N[S]\} 
 \\
& &0\le x_v\le 1, ~~    \forall  v\in V-S 
\end{eqnarray}

Next, let $A=S, B=E(S)$ and $C=V-(A\cup B)$, and  apply the rounding scheme in algorithms $A_1$ or $A_2$ to $C$,  and let $H$ and $U$ be corresponding sets, and output the set  $S\cup H\cup U$. 
\end{tcolorbox}

\begin{rem}\label{r2} 
{\sl  Note that by Theorem \ref{t1} Algorithm $H$ can be implemented in polynomial time. Furthermore, 
$|S\cup H\cup U|\le {\alpha.d}+3a(G)N^*\le \alpha.(ln(\Delta)+1)+3a(G)).\gamma(G)$, and thus  Algorithm  $H$ has a bounded performance ratio. }

\end{rem}



\section{Performance on Planar Graphs, k-Planar Graphs, and k-Trees}

In this section, we compare the performance ratios of Greedy, $A_1$, $A_2$, $A_1$ Hybrid, and $A_2$ Hybrid on planar graphs, k-planar graphs k-trees. In Tables 2 and 3, we present the performance of the algorithms on k-trees where $k=\lfloor |V|^{0.25}\rfloor$ and k-planar graphs where $k=\lfloor \ln{(|V|)}\rfloor$, respectively. These graphs are dense. We also present the algorithms' performance on sparse k-trees and sparse k-planar graphs in tables 4 and 5. The planar graphs k-trees, and k-planar graphs were all made using $\dagger$ described in section 1.5. \newline

In most cases, the $A_{1}$ and $A_{2}$ variants of the hybrid algorithm outperformed the others, producing the lowest performance ratio to the LP lower bound $L^*$. Greedy performs close to hybrid and outperforms it for the larger dense k-trees and a few of the k-planar graphs. The LP-rounding algorithms performed the worst across the board. All algorithms were able to compute dominating sets in less than 2 seconds across the different types of graphs and their range of sizes.\\

The arboricity of each of the planar graphs is at most 3. For k-trees, we use $\lceil k - \frac{(k/2)(k-1)}{N-1} \rceil$ for arboricity. For k-planar graphs, we use the upper bound of  $\lceil 8\sqrt{k}\rceil$ on arboricity.

\begin{table}[h!]\centering
\caption{Results for Planar Graphs}\label{tab: }
\scriptsize
\begin{tabular}{lrrrrrr}\toprule
$n, m$ & $L^*$ & Greedy/$L^*$ & $A_1$/$L^*$ & $A_1$ Hybrid/$L^*$ &$A_2$/$L^*$ & $A_2$ Hybrid/$L^*$ \\\midrule
2000, 5980 &316.93 &1.12 &1.40 &1.11 &1.39 &1.11 \\
4000, 11972 &620.72 &1.16 &1.35 &1.14 &1.34 &1.14 \\
6000, 17978 &942.59 &1.13 &1.29 &1.13 &1.29 &1.13 \\
8000, 23974 &1239.16 &1.14 &1.41 &1.13 &1.40 &1.13 \\
10000, 29972 &1579.06 &1.13 &1.27 &1.13 &1.27 &1.13 \\
12000, 35973 &1874.66 &1.13 &1.36 &1.12 &1.35 &1.12 \\
14000, 41974 &2185.35 &1.14 &1.33 &1.14 &1.32 &1.14 \\
16000, 47975 &2514.62 &1.14 &1.33 &1.13 &1.33 &1.13 \\
18000, 53971 &2811.98 &1.15 &1.35 &1.14 &1.35 &1.14 \\
20000, 59971 &3127.20 &1.14 &1.32 &1.13 &1.31 &1.13 \\
\bottomrule
\end{tabular}
\end{table}

\begin{table}[!htbp]\centering
\caption{Results for k-Trees where $k=\lfloor |V|^{0.25}\rfloor$}\label{tab: }
\scriptsize
\begin{tabular}{lrrrrrr}\toprule
$n, m$ &$L^*$ &Greedy/$L^*$ &$A_1$/$L^*$ & $A_1$ Hybrid/$L^*$ &$A_2$/$L^*$ & $A_2$ Hybrid/$L^*$ \\\midrule
2000, 13972 &15.00 &1.07 &1.20 &1.00 &1.20 &1.00 \\
4000, 31964 &10.00 &1.00 &1.00 &1.00 &1.00 &1.00 \\
6000, 53955 &11.00 &1.00 &1.00 &1.00 &1.00 &1.00 \\
8000, 71955 &13.00 &1.00 &1.00 &1.00 &1.00 &1.00 \\
10000, 99945 &11.19 &1.07 &2.23 &1.07 &2.23 &1.07 \\
12000, 119945 &12.00 &1.00 &1.00 &1.00 &1.00 &1.00 \\
14000, 139945 &18.50 &1.08 &1.89 &1.14 &1.89 &1.14 \\
16000, 175934 &11.25 &1.16 &1.60 &1.33 &1.60 &1.33 \\
18000, 197934 &11.00 &1.18 &2.00 &1.18 &2.00 &1.18 \\
20000, 219934 &10.50 &1.14 &1.43 &1.43 &1.43 &1.43 \\
\bottomrule
\end{tabular}
\end{table}

\begin{table}[h!]\centering
\caption{Results for k-Planar Graphs where $k=\lfloor \ln{(|V|)}\rfloor$}\label{tab: }
\scriptsize
\begin{tabular}{lrrrrrr}\toprule
$n, m$ &$L^*$ &Greedy/$L^*$ &$A_1$/$L^*$ & $A_1$ Hybrid/$L^*$ &$A_2$/$L^*$ & $A_2$ Hybrid/$L^*$ \\\midrule
2000, 12986 &151.97 &1.26 &2.16 &1.24 &2.11 &1.24 \\
4000, 27254 &289.69 &1.27 &2.65 &1.29 &2.64 &1.29 \\
6000, 40885 &431.77 &1.26 &2.50 &1.26 &2.50 &1.26 \\
8000, 54568 &568.01 &1.24 &2.57 &1.25 &2.57 &1.25 \\
10000, 71414 &684.20 &1.27 &2.57 &1.28 &2.56 &1.28 \\
12000, 85580 &821.65 &1.26 &2.62 &1.27 &2.62 &1.27 \\
14000, 100241 &957.77 &1.25 &2.47 &1.26 &2.46 &1.26 \\
16000, 114270 &1098.18 &1.27 &2.21 &1.27 &2.21 &1.27 \\
18000, 128725 &1238.09 &1.27 &2.23 &1.27 &2.22 &1.27 \\
20000, 142891 &1368.44 &1.26 &2.24 &1.25 &2.23 &1.25 \\
\bottomrule
\end{tabular}
\end{table}

\begin{table}[!htp]\centering
\caption{Results for k-Trees where $k=5$}\label{tab: }
\scriptsize
\begin{tabular}{lrrrrrrr}\toprule
$n, m$ &$L^*$ &Greedy/$L^*$ &$A_1$/$L^*$ & $A_1$ Hybrid/$L^*$ &$A_2$/$L^*$ &$A_2$ Hybrid/$L^*$ \\\midrule
2000, 9985 &39.00 &1.05 &1.08 &1.05 &1.08 &1.05 \\
4000, 19985 &70.50 &1.04 &1.06 &1.04 &1.06 &1.04 \\
6000, 29985 &90.83 &1.03 &1.17 &1.03 &1.17 &1.03 \\
8000, 39985 &132.25 &1.03 &1.07 &1.03 &1.07 &1.03 \\
10000, 49985 &158.00 &1.03 &1.03 &1.03 &1.03 &1.03 \\
12000, 59985 &209.67 &1.02 &1.08 &1.02 &1.08 &1.02 \\
14000, 69985 &225.58 &1.04 &1.09 &1.04 &1.09 &1.04 \\
16000, 79985 &270.25 &1.02 &1.09 &1.02 &1.09 &1.02 \\
18000, 89985 &291.83 &1.02 &1.06 &1.02 &1.06 &1.02 \\
20000, 99985 &339.58 &1.04 &1.08 &1.04 &1.08 &1.04 \\
\bottomrule
\end{tabular}
\end{table}

\begin{table}[!htp]\centering
\caption{Results for k-Planar Graphs where $k=5$}\label{tab: }
\scriptsize
\begin{tabular}{lrrrrrrr}\toprule
$n, m$ &$L^*$ &Greedy/$L^*$ &$A_1$/$L^*$ & $A_1$ Hybrid/$L^*$ &$A_2$/$L^*$ & $A_2$ Hybrid/$L^*$ \\\midrule
2000, 11465 &171.42 &1.19 &1.65 &1.20 &1.65 &1.20 \\
4000, 23033 &336.57 &1.21 &1.63 &1.22 &1.63 &1.22 \\
6000, 34577 &510.02 &1.24 &2.20 &1.25 &2.19 &1.25 \\
8000, 46130 &680.88 &1.25 &1.91 &1.25 &1.91 &1.25 \\
10000, 57786 &840.92 &1.23 &2.12 &1.24 &2.10 &1.24 \\  
12000, 69220 &1019.54 &1.23 &2.02 &1.22 &2.02 &1.22 \\
14000, 80680 &1181.05 &1.22 &1.90 &1.22 &1.90 &1.22 \\
16000, 92300 &1355.13 &1.23 &2.03 &1.23 &2.03 &1.23 \\
18000, 103862 &1516.14 &1.24 &1.99 &1.24 &1.99 &1.24 \\
20000, 115354 &1689.35 &1.22 &2.08 &1.21 &2.08 &1.21 \\
\bottomrule
\end{tabular}
\end{table}

\vspace{3cm}
\section{Performance on Hypercubes and k-Queen Graphs}

In this section, we present the performance of Greedy, $A_1$, $A_2$, $A_1$ Hybrid, and $A_2$ Hybrid on hypercubes from 5-12 dimensions and k-Queens graphs.\\

Table 6 compares the performance ratios of the algorithms on hypercubes. We use the arboricity for hypercubes $a=\lfloor{d/2}+1\rfloor$ for LP rounding and hybrid \cite{KM}. For k-Queens graphs, arboricity is unknown, so we use the upper  bound $3(k-1)$, where $k$ is the length of the chessboard. \\

For both hypercubes and k-Queens graphs, Greedy performs the best, followed by $A_1$ Hybrid and $A_2$ Hybrid. $A_1$ and $A_2$ LP rounding perform the worst by far. This is not surprising as LP Rounding approaches are known to in general perform worse on dense graphs than sparse graphs. Solutions were computed in under 8 seconds for all graphs and algorithms.

\begin{adjustwidth}{-2.5 cm}{-2.5 cm}\centering\begin{threeparttable}[!htb]\centering
\caption{Results for Hypercubes}\label{tab:}
\scriptsize
\begin{tabular}{lrrrrrr}\toprule
$n, m$ &$L^*$ &Greedy/$L^*$ &$A_1$/$L^*$ & $A_1$ Hybrid/$L^*$ &$A_2$/$L^*$ & $A_2$ Hybrid/$L^*$ \\\midrule
5, 80 &5.33 &1.50 &3.00 &1.50 &3.00 &1.50 \\
6, 192 &9.14 &1.75 &7.00 &1.75 &7.00 &1.75 \\
7, 448 &16.00 &1.00 &1.00 &1.00 &1.00 &1.00 \\
8, 1024 &28.44 &1.13 &9.00 &1.13 &9.00 &1.13 \\
9, 2304 &51.20 &1.25 &7.07 &2.99 &7.07 &2.99 \\
10, 5120 &93.09 &1.38 &11.00 &2.70 &11.00 &2.70 \\
11, 11264 &170.67 &1.50 &6.59 &2.85 &6.59 &2.85 \\
12, 24576 &315.08 &1.63 &13.00 &3.14 &13.00 &3.14 \\

\bottomrule
\end{tabular}
\end{threeparttable}\end{adjustwidth}

\begin{adjustwidth}{-2.5 cm}{-2.5 cm}\centering\begin{threeparttable}[!htb]\centering
\caption{Results for k-Queens Graphs}\label{tab: }
\scriptsize
\begin{tabular}{lrrrrrr}\toprule
$n, m$  &$L^*$ &Greedy/$L^*$ &$A_1$/$L^*$ & $A_1$ Hybrid/$L^*$ &$A_2$/$L^*$ & $A_2$ Hybrid/$L^*$ \\\midrule
225, 5180 &4.89 &2.05 &38.45 &6.75 &36.40 &6.75 \\
256, 6320 &5.19 &1.93 &46.98 &7.70 &43.90 &7.12 \\
289, 7616 &5.50 &1.82 &45.84 &8.91 &44.03 &8.91 \\
324, 9078 &5.80 &1.90 &50.34 &9.83 &48.27 &9.83 \\
361, 10716 &6.10 &1.97 &52.42 &9.67 &50.78 &9.67 \\
400, 12540 &6.41 &2.03 &56.81 &10.14 &53.06 &9.68 \\
441, 14560 &6.71 &1.94 &59.89 &11.32 &56.91 &11.17 \\
484, 16786 &7.02 &2.00 &63.86 &9.55 &59.29 &9.12 \\
529, 19228 &7.32 &1.91 &65.83 &10.38 &62.83 &10.11 \\
576, 21896 &7.62 &1.97 &70.82 &11.93 &64.00 &11.67 \\
625, 24800 &7.93 &2.02 &74.15 &10.47 &69.61 &10.34 \\
676, 27950 &8.23 &1.94 &76.27 &11.78 &68.50 &11.30 \\
729, 31356 &8.54 &1.87 &80.80 &11.83 &74.48 &11.13 \\
784, 35028 &8.84 &1.92 &80.07 &14.82 &74.64 &14.25 \\
841, 38976 &9.15 &1.97 &85.81 &12.02 &78.81 &11.70 \\
900, 43210 &9.45 &2.01 &87.18 &12.91 &81.26 &12.38 \\

\bottomrule
\end{tabular}
\end{threeparttable}\end{adjustwidth}

\section{Performance on Real-World Graphs}

In this section, we present the performance of LP rounding, greedy, and hybrid on the real-world social network graphs from Google+ \cite{Ch}, Pokec \cite{Ch}, and DIMACS \cite{Bader}. Each of these graphs are sparse, but their arboricity is unknown. Since arboricity is unknown, we experiment with the threshold applied during LP rounding and hybrid, starting with $1/3a'$, where $a'=\lceil |E| / (|V| - 1)\rceil$ is  a  lower bound on arborictiy . We call LP Rounding with this threshold Algorithm  $A_{1}'$. Similarly,  Algorithm $A_{2}'$ has threshold $1/2a'+1$. Through experimentation, the best threshold which we found was $2/a'$; the resulting Algorithm is called  $A_{3}$. \newline

In Table 8, we compare the solution size of $A_{1}'$, $A_{2}'$, and $A_{3}$, along with their hybrid analogs and greedy, to the LP lower bound $L^*$ on the Google+ graphs. Table 9 compares the same algorithms on the Pokec graphs. In Table 10, we compare the performance ratio to the LP lower bound for these algorithms on 3 social network graphs from DIMACS. In Tables 8, 9 and 10, LP Rounding performs better than the greedy and hybrid approaches, with greedy being the worst out of the algorithms tested. Out of the LP rounding approaches, $A_3$ performs the best. \newline 

\begin{adjustwidth}{-2.5 cm}{-2.5 cm}\centering\begin{threeparttable}[!htb]
\caption{Results for \emph{Google}+ Graphs}\label{tab: }
\scriptsize
\begin{tabular}{lrrrrrrrr}\toprule
$n, m$ &$L^*$ &Greedy & $A_1'$ & $A_2'$ & $A_3$ &$A_1'$ Hybrid & $A_2'$ Hybrid & $A_3$ Hybrid \\\midrule
500, 1006 &42 &42 &42 &42 &42 &42 &42 &42 \\
2000, 5343 &170 &176 &170 &170 &170 &176 &176 &176 \\
10000, 33954 &860 &900 &864 &864 &864 &893 &893 &893 \\
20000, 81352 &1715 &1817 &1730 &1730 &1716 &1800 &1800 &1800 \\
50000, 231583 &4565 &4849 &4651 &4607 &4585 &4790 &4790 &4790 \\
\bottomrule
\end{tabular}
\end{threeparttable}\end{adjustwidth}

\begin{adjustwidth}{-2.5 cm}{-2.5 cm}\centering\begin{threeparttable}[!htb]
{\caption{Results for \emph{Pokec} Graphs}}\label{tab: }
\scriptsize
\begin{tabular}{lrrrrrrrr}\toprule
$n, m$ &$L^*$ &Greedy & $A_1'$ & $A_2'$ &$A_3$ &$A_1'$ Hybrid & $A_2'$ Hybrid &$A_3$ Hybrid \\\midrule
500, 993 &16 &16 &16 &16 &16 &16 &16 &16 \\
2000, 5893 &75 &75 &75 &75 &75 &75 &75 &75 \\
10000, 44745 &413 &413 &413 &413 &413 &413 &413 &413 \\
20000, 102826 &921 &928 &921 &921 &921 &923 &923 &923 \\
50000, 281726 &2706 &2773 &2712 &2712 &2712 &2757 &2757 &2743 \\
\bottomrule
\end{tabular}
\end{threeparttable}\end{adjustwidth}

\vspace{0.5 cm}

Compared to the best results from \cite{Ch}, which used a randomized local search algorithm that is run for up to one hour, LP Rounding approaches generally produced a smaller or as good solution using significantly less run-time at less than 0.5 seconds for each graph.

\begin{adjustwidth}{-2.5 cm}{-2.5 cm}\centering\begin{threeparttable}[!htb]\centering
\caption{Results for DIMACS Graphs}\label{tab: }
\scriptsize
\begin{tabular}{llrrrrrrrr}\toprule
Graph &$n, m$ &$L^*$ &Greedy/$L^*$ &$A_1'$/$L^*$ & $A_1'$ Hybrid$/L^*$ &$A_2'$/$L^*$ & $A_2'$ Hybrid/$L^*$ & $A_3/L^*$ & $A_3$ Hybrid$/L^*$ \\\midrule
coAuthorsDBLP &299067, 977676 &43969.00 &1.02 &1.00 &1.02 &1.00 &1.02 & 1.00 & 1.02 \\
coPapersCiteseer &434102, 16036720 &26040.92 &1.12 &1.01 &1.12 &1.01 &1.12 & 1.01 & 1.12\\
citatinCiteseer &268495, 1156647 &43318.85 &1.04 &1.03 &1.04 &1.03 &1.04 & 1.02 & 1.04 \\
\bottomrule
\end{tabular}
\end{threeparttable}\end{adjustwidth}

\vspace{0.5 cm}

Table 11 shows an example of a 7 million+ vertices graph where $A_1$ and $A_2$ cannot be run as a result of the large size. For hybrid approaches, using the first $d/2$ vertices from the greedy solution, where $d$ is the size of the greedy solution, resulted in the use of too much memory. We instead used the first $3d/4$ vertices from the greedy solution. Both $A_1$ Hybrid and $A_2$ Hybrid performed better than greedy. Greedy took 14 seconds to produce a solution while hybrid took 107 seconds. $max\{M^*, N^*\}$ is provided as a lower bound on $L^*$, and therefore, $\gamma(G)$.

\begin{adjustwidth}{-2.5 cm}{-2.5 cm}\centering\begin{threeparttable}[!htb]
\caption{Results for the DIMACS Great Britain Street Network}\label{tab: }
\scriptsize
\begin{tabular}{lrrrrrr}\toprule
$n, m$ &$M^*$ &$N^*$ &max\{$M^*, N^*$\} &Greedy & $A_1$ Hybrid & $A_2$ Hybrid \\\midrule
7733822, 8156517 &1314133 &1357189 &1357189 &2732935 &2724608 &2724608 \\
\bottomrule
\end{tabular}
\end{threeparttable}\end{adjustwidth}

\vspace{0.5 cm}

\pagebreak


\begin{thebibliography}{}

\bibitem{BM}T. Bohme and B. Mohar, Domination, packing and excluded minors, Electronic Journal of Combinatorics, 10 (2003), p. N9.

\bibitem {Bader} David A. Bader, Andrea Kappes, Henning Meyerhenke, Peter Sanders, Christian Schulz and Dorothea Wagner. Benchmarking for Graph Clustering and Partitioning. In Encyclopedia of Social Network Analysis and Mining, pages 73-82. Springer, 2014.

\bibitem {Ban}  Bansal and S. W. Umboh. Tight approximation bounds for dominating set on graphs of bounded arboricity. Information Processing Letters. (2017), 21-24. 

\bibitem{Be} Claude Berge, Two Theorems in Graph Theory. Proc National Acad Sci U S A. 1957 Sep 15; 43(9): 842–844.

\bibitem{Bert} R. Bertolo \& P. R. J. Östergård \& W. D. Weakley. An Updated Table of Binary/Ternary Mixed Covering Codes. Journal of Combinatorial Designs. 12. 157 - 176. 2004. 

\bibitem{BG} H. Brönnimann, M. T. Goodrich, 
 Almost Optimal Set Covers in Finite VC-Dimension
Discrete and Computational Geometry 14(1):463-479, 1995.

\bibitem{Bur} A.p. Burger and C.m. Mynhardt. An Upper Bound for the Minimum Number of Queens Covering the nxn Chessboard. Discrete Applied Mathematics, vol. 121, no. 1-3, pp. 51–60., 2002

\bibitem  {Cam} A Campan, T. M. Truta, and M. Beckerich. Fast Dominating Set Algorithms for Social Networks,  MAICS. (2015). 


\bibitem {Ch} D. Chalupa, An Order-based Algorithm for Minimum Dominating Set with Application in Graph Mining, Information Sciences (2018), 101-116.

\bibitem {Chv} V. Chvatal. A greedy heuristic for the set-covering problem, Mathematics of Operations Research, 4(3):233–235, 1979. 

\bibitem{CLR} T.  H. Cormen, C. E. Leiserson, R. L. Rivest, C. Stein. Introduction to Algorithms. MIT Press, 1990. 

\bibitem{CL} G.  Chartrand, L.  Lesniak, P.  Zhang, Graphs and Digraphs, CRC Press, 2010.
 
\bibitem{CC} B. N. Clark, C. J. Colbourn, and D. S. Johnson. Unit disk graphs. Discrete Mathematics, 86(1-3):165–177, 1990.

\bibitem{AS} {Alexander Schrijver}, Theory of Linear and Integer Programming
\bibitem{D1} Z. Dvořák , Constant-factor approximation of domination number in sparse graphs, European Journal of Combinatorics, 34 (2013), 833–840

\bibitem {D2} Z. Dvořák,  On distance r-dominating and 2r-independent sets in sparse graphs,  Journal of Graph Theory, (2017). 
\bibitem{DD}  I. Dinur, D.  Steurer, Analytical approach to parallel repetition, in: Symposium on Theory of Computing, STOC, 2014, 624–633.

\bibitem {Eu} S. Eubank. V.S., Anil Kumar, M. V. Marathe, A. Srinivasan, and N. Wang. Structural and Algorithmic Aspects of Massive Social Networks SODA (2004), 718-727.

\bibitem{Ga} H. N Gabow, H. H. Westermann, Forests, frames, and games: Algorithms for matroid sums and applications. Algorithmica. 7 (1): 465–497, 1992.

\bibitem {JHC} H. Jung, M. K. Hasan, and K. Chwa,  Improved Primal-Dual Approximation Algorithm for the Connected Facility Location Problem. (2008), 265-277.

\bibitem{KM}N. Karisani, E.S. Mahmoodian, On the Construction of Tree Decompositions of Hypercubes. 2013. 

\bibitem{HHS}T. W. Haynes, S. Hedetniemi, P. Slater.
Fundamentals of Domination in Graphs, CRC press, 1988.

\bibitem{Karp} R.  M. Karp , Reducibility Among Combinatorial Problems, . In R. E. Miller; J. W. Thatcher; J.D. Bohlinger (eds.). Complexity of Computer Computations. New York: Plenum, (1972),  pp. 85–103.

\bibitem{Kha} L. g. Khachiyan Polynomial Algorithms in Linear Programming. USSR Computational Mathematics and Mathematical Physics, vol. 20, no. 1, pp. 53–72., 1980.

\bibitem{Lo}  L.  Lovasz, On the Ratio of Optimal Integral and Fractional Covers. Discrete Mathematics, Vol. 13, 1975, 383–390.

\bibitem {Par} A. K. Parekh. Analysis of a Greedy Heuristic For Finding Small Dominating Sets in Graphs,  Information Processing Letters. (1991), 237-240.

\bibitem {Sa} L. A. Sanchis, Experimental Analysis of Heuristic Algorithms for the Dominating Set Problem. Algorithmica, (2002), 3-18.


\bibitem {Si} S. Siebertz,  Greedy domination on biclique-free graphs. Information Processing Letters. (2019), 64-67

\bibitem {Yang} J. Yang and J. Leskovec. Defining and Evaluating Network Communities based on Ground-truth. International Conference on Data Mining. (2012), 745-754.

\bibitem{Nash} C. St. J. A. Nash-Williams  Decomposition of finite graphs into forests,  Journal of the London Mathematical Society. 39 (1): 12, 1964.







 
\end{thebibliography}
\end{document}